\documentclass[12pt]{article}
\pdfoutput=1
\usepackage{jheppub}
\usepackage{verbatim}

\newcommand{\be}{\begin{equation}}
\newcommand{\ee}{\end{equation}}
\newcommand{\bfig}{\begin{figure}\begin{center}}
\newcommand{\efig}{\end{center}\end{figure}}
\newcommand{\bi}{\begin{itemize}}
\newcommand{\ei}{\end{itemize}}

\newcommand{\lan}{\langle}
\newcommand{\ran}{\rangle}

\newcommand{\cpn}{$\mathbb{CP}^{N-1}$}
\begin{document}
%\subheader{empty}
\title{Wormholes, Emergent Gauge Fields, and the Weak Gravity Conjecture}
\author{Daniel Harlow}
\affiliation{Center for the Fundamental Laws of Nature, Harvard University, Cambridge MA, 02138 USA}
%\affiliation[b]{Princeton Center for Theoretical Science, Princeton University, Princeton NJ 08540 USA}

\emailAdd{dharlow@fas.harvard.edu}
\abstract{This paper revisits the question of reconstructing bulk gauge fields as boundary operators in AdS/CFT.  In the presence of the wormhole dual to the thermofield double state of two CFTs, the existence of bulk gauge fields is in some tension with the microscopic tensor factorization of the Hilbert space.  I explain how this tension can be resolved by splitting the gauge field into charged constituents, and I argue that this leads to a new argument for the ``principle of completeness'', which states that the charge lattice of a gauge theory coupled to gravity must be fully populated.  I also claim that it leads to a new motivation for (and a clarification of) the  ``weak gravity conjecture'', which I interpret as a strengthening of this principle.  This setup gives a simple example of a situation where describing low-energy bulk physics in CFT language requires knowledge of high-energy bulk physics.  This contradicts to some extent the notion of ``effective conformal field theory'', but in fact is an expected feature of the resolution of the black hole information problem.  An analogous factorization issue exists also for the gravitational field, and I comment on several of its implications for reconstructing black hole interiors and the emergence of spacetime more generally.}
\maketitle

\section{Introduction}
The subject of reconstruction of bulk fields as boundary operators in AdS/CFT has generated a lot of interest in the last few years \cite{Banks:1998dd,Hamilton:2006az,Kabat:2011rz,Heemskerk:2012mn,Morrison:2014jha}.  In particular the additional subtleties present when there are gauge symmetries in the bulk has recently received a fair bit of attention \cite{Kabat:2012hp,Kabat:2012av,Heemskerk:2012np,Papadodimas:2013jku,Harlow:2014yoa,Kabat:2013wga,Almheiri:2014lwa,Giddings:2015lla,Donnelly:2015hta,Donnelly:2015taa}.  Since the graviton will always be present in any example of AdS/CFT, it is clear that a complete understanding of these issues is necessary for a full understanding of the correspondence.  In \cite{Harlow:2014yoa} a new puzzle in the CFT reconstruction of bulk fields in the presence of gauge symmetries was pointed out (see also \cite{Engelhardt:2015fwa} for some related discussion); the goal of this paper is to describe this puzzle in more detail and propose a resolution for it.  I will focus predominantly on the simple case of a single $U(1)$ gauge symmetry associated with a vector potential $A_\mu$ in the bulk. For obvious reasons I will refer to this as the electromagnetic case, and will use the standard terminology from electromagnetism to describe it.  

\bfig
\includegraphics[height=5cm]{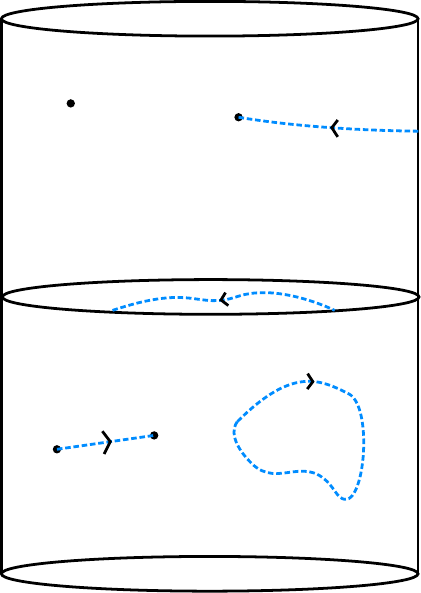}\hspace{3cm}\includegraphics[height=5cm]{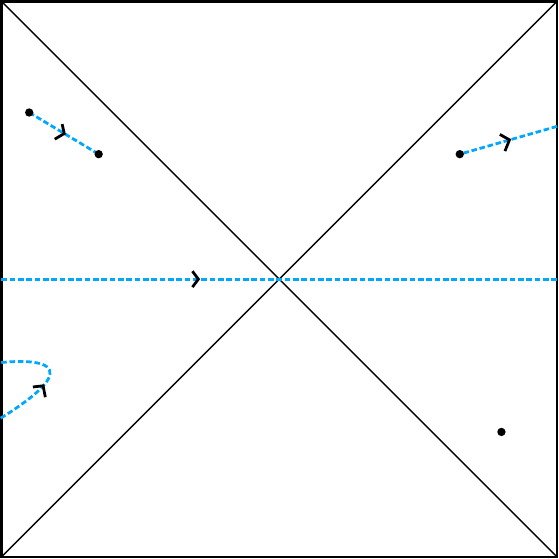}
\caption{Examples of gauge-invariant operators for electromagnetism on the AdS and AdS-Schwarzschild backgrounds.  The directed blue dashed lines represent Wilson lines, and dots with Wilson lines ending (beginning) on them represent local operators creating positive (negative) charge.  Wilson lines can also end on the boundary, or be closed into loops.  Note that in the AdS-Schwarzschild background there is a new kind of Wilson line, that stretches from one boundary to the other.}\label{giopsfig}
\efig
In the vicinity of the vacuum in Hilbert space, there is a rather complete and intuitive understanding of how to reconstruct bulk gauge-invariant operators in the CFT in the electromagnetic case \cite{Kabat:2012hp,Kabat:2012av,Heemskerk:2012np}.\footnote{By ``gauge-invariant'' here, I mean gauge-invariant under the $U(1)$ gauge symmetry of electromagnetism.  Additional work is needed to define operators that are also gauge-invariant under gravitational diffeomorphisms.  For now I will ignore this issue, with the excuse that at leading order in the $1/N$ expansion it does not arise since we are just doing quantum field theory in a fixed background.  This is  not totally satisfactory, after all avoiding large gravitational backreaction is important in defining what ``in the vicinity of the vacuum'' really means \cite{Almheiri:2014lwa}.  I will return to this issue in section \ref{gravsec} below.}  These consist of local uncharged operators such as the field strength or the electromagnetic current, as well as nonlocal operators such as closed Wilson lines or Wilson lines ending on charged operators or the boundary.\footnote{Wilson lines are allowed to end on the boundary since, in the standard quantization of the gauge field in an asymptotically AdS background, allowed gauge transformations must vanish at infinity.  I review this somewhat in appendix \ref{gaugeapp}, see for example \cite{Marolf:2006nd} for a detailed discussion.}  The left diagram of figure \ref{giopsfig} illustrates a representative collection of such operators.\footnote{Strictly localized Wilson line operators are somewhat singular, since they create an infinitely thin line of electric flux that has infinite energy.  They can be regulated by smearing them against some sort of transverse profile; I will not be explicit about it in this paper but one should always remember that this smearing is implicitly present.}  
Some details of this construction are reviewed in appendix \ref{gaugeapp} for completeness, but they will not really be important for this paper since we will mostly rely on bulk techniques.  The only important point is that all of the operators in this diagram have CFT representations in terms of the global current dual to the gauge field and the local operators dual to any charged bulk fields.  

The situation becomes more interesting when we consider other backgrounds, such as the AdS-Schwarzschild geometry shown in the right side of figure \ref{giopsfig}.  From the figure it is clear that in such a background there is a new kind of gauge-invariant operator: a Wilson line that runs through the wormhole from one boundary to another.  Acting with such an operator on the Hartle-Hawking state, which above the Hawking-Page transition is widely expected to be dual to the thermofield double state of the two CFTs \cite{Maldacena:2001kr}, one creates electric flux through the wormhole.  In other words this action makes the wormhole ``a little bit Reissner-Nordstrom''.    This Wilson line has the interesting property that it \textit{only exists} in states where the two boundaries are connected by a wormhole; its existence is thus in some sense a diagnostic for whether or not the two sides are connected through the bulk \cite{Harlow:2014yoa,Engelhardt:2015fwa}.  It seems likely that if we understand how to describe it in the CFTs, we will have learned something about how the wormhole arises in the appropriate states.\footnote{In figure \ref{giopsfig} I did not draw any operators in the future or past interior regions; in particular I took the wormhole-threading Wilson line to puncture the bifurcate horizon.  Once we have the full set of exterior operators together with these Wilson lines, the interior operators can be obtained by bulk Cauchy evolution \cite{Heemskerk:2012mn}.  This evolution becomes problematic if we consider interior operators for observers who jump in at very early or very late times, see eg \cite{Almheiri:2013hfa,Papadodimas:2015xma}, but as long as we focus on experiments that do not involve very long time separations on the boundary, we can just use the symmetry of the state to move everything to the vicinity of $t=0$.}  

It is not so clear that such wormhole-threading Wilson lines can be reconstructed as CFT operators; in particular the direct method of \cite{Kabat:2012hp,Kabat:2012av,Heemskerk:2012np} based on using the CFT current will not work. The problem is that any attempt to cut a Wilson line into parts involving only the gauge field leads to operators which are not separately gauge-invariant.  This is puzzling however, since the microscopic Hilbert space of the theory is a tensor product of the two boundary CFTs.  This means that \textit{any} two-CFT operator can be decomposed into a sum of tensor products of one-CFT operators.  These one-CFT operators are themselves gauge-invariant, since any operator in the CFT is by definition invariant under bulk gauge transformations.\footnote{If this statement is not obvious, recall that in any gauge theory we define the set of physical states as those which are gauge-invariant.  But in the CFT all states are physical.  Any operator that acts on them must thus be gauge-invariant.}  So how are the CFTs cutting the Wilson line in the bulk? In what follows, I will refer to this question as the \textit{factorization problem}.

The statement that the two-CFT Hilbert space factorizes despite the existence of wormhole-threading Wilson lines is very powerful, for example it immediately implies that the CFT must have local operators (or equivalently states) that are charged under the symmetry generated by the current that is dual to the gauge field.  In the bulk this is a standard piece of lore; any theory with an Einstein-Maxwell sector must also have objects that are charged under the gauge field \cite{Polchinski:2003bq,Banks:2010zn}.  Indeed recall that if we define charge operators $Q_R$ and $Q_L$ in the right and left CFTs respectively, then the thermofield double is annihilated by $Q_R+Q_L$ (I define these operators, and the thermofield double, more carefully in appendix \ref{gaugeapp} below).  The operator $Q_L-Q_R$ does not annihilate the thermofield double, but it does still have zero expectation value, since $Q_R$ and $Q_L$ do separately by the CPT-invariance of the thermal state.  If we now act on the thermofield double with a Wilson line in the charge $n$ representation threading the wormhole from left to right, we get a state in which the expectation value of $Q_L-Q_R$ is $2n$.  For $n\neq 0$ such a state clearly can only exist if the single-CFT charge operator $Q$ is nontrivial; charged states must exist!  The Hilbert space factorization is essential to this argument, in section \ref{facsec} below we will see an example of how it could fail otherwise.\footnote{One might try to derive this conclusion from abstract conformal field theory, but so far there does not seem to be a proof outside of two dimensions that the existence of a current requires operators that are charged under that current. I thank Thomas Dumitrescu and Sasha Zhiboedov for discussions of this point.}

As we proceed it will become clear that the resolution of the factorization problem depends on short-distance physics in the bulk. One indication of this, which I will review in section \ref{facsec} below, is that in Hamiltonian lattice gauge theory \cite{Kogut:1974ag} there is no gauge-invariant way whatsoever to cut a Wilson line in the fundamental representation. The microscopic Hilbert space therefore does \textit{not} tensor factorize into separate spatial regions \cite{Donnelly:2011hn,Casini:2013rba,Radicevic:2014kqa,Huang:2014pfa,Radicevic:2015sza}.  We thus are really asking what kind of high-energy cutoff AdS/CFT provides for bulk electromagnetism; apparently it must be of a different sort than that provided by lattice gauge theory.  

In this paper I will not be able to address this question in full, after all doing so would essentially amount to constructing bulk quantum gravity.  My goal will instead be to give a rather general mechanism whereby the factorization problem can be resolved entirely within bulk effective field theory.  The mechanism is to cut the Wilson line by replacing a segment by a pair of oppositely charged fields at the ends of the segment, as illustrated in figure \ref{cutfig}.  
\bfig
\includegraphics[height=2cm]{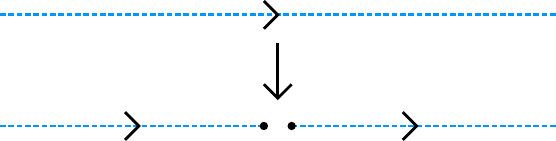}
\caption{Cutting a Wilson line by a pair of oppositely charged fields. I will argue below that there is a natural ``gauge-covariant operator product expansion'' that ensures these operators are proportional to each other in low-energy correlation functions.  Said differently, the bottom operator flows to the top one under the renormalization group.  In the case of a wormhole-threading Wilson line, the lower operator has a simple CFT description since we can represent the left half in the left CFT and the right half in the right CFT.}\label{cutfig}
\efig
In order for this type of replacement to reproduce the Wilson line at all scales where it exists, we will in fact need the gauge field to emerge from a microscopic theory with only the charges.  There are many examples of theories which achieve this in some sort of large $N$ limit \cite{Lee:2010fy}, and I will revisit one of them, the $\mathbb{CP}^{N-1}$ nonlinear-$\sigma$ model in $d\geq 2$ \cite{D'Adda:1978uc,Witten:1978bc,Polyakov:1987ez}, in detail to see how it explicitly resolves the factorization problem.\footnote{This proposal seems closely related to the discussion of \cite{Donnelly:2015hxa}, where it is argued that fictitious charges near the horizon are necessary to correctly match Euclidean and Lorentzian calculations of black hole entropy in the presence of a Maxwell field.  One can interpret this paper as claiming that those charges aren't actually so fictitious.}   

I do not mean to argue that this mechanism is \textit{necessarily} how the factorization problem is resolved in all examples of AdS/CFT; indeed there is an analogous factorization problem for gravity, whose resolution is almost certainly \textit{not} described by effective field theory.  The hope is rather that the basic features of the mechanism hold more generally; indeed many of them are already known or expected to be true in string theory for independent reasons. That they follow so naturally from this idea suggests, at least to my mind, that the truth in general can't be too different. 
The main lessons are:\footnote{In this paper the word ``fundamental'' always means ``transforming in the minimal positive charge representation''.  It is never used in the colloquial sense of ``fundamental field''.  When the latter meaning is intended, I will use the terms ``microscopic'' or ``short-distance''.} 
\bi
\item Cutting a Wilson line in the fundamental representation requires the existence of bulk fields of fundamental charge.
\item These charged fields can have rather large masses, in which case describing the ``low-energy'' Wilson line in the CFT requires fairly high-dimension CFT operators.
\item They cannot however be so heavy that they are black holes.  In fact, obtaining a weak gauge coupling in the bulk requires the charges to be parametrically lighter than the Planck scale.  The mechanism thus seems to require the theory to obey the ``weak gravity conjecture'' of \cite{ArkaniHamed:2006dz}, in the form that demands the existence a fundamentally-charged particle of mass $m$, such that $m \sqrt{G}\lesssim q$, where $q$ is the gauge coupling.
\ei
All three points are closely related to prior discussions.  The required existence of fundamental charges is a long-standing piece of lore in quantum gravity, it is sometimes called the ``principle of completeness'' \cite{Polchinski:2003bq,Banks:2010zn}.  It certainly seems to be true in string theory, but so far to my knowledge there has not really been a convincing general argument given for it.\footnote{In fact the general argument given above for the nonvanishing of $Q$, together with the single-CFT vanishing of the expectation value of $Q$ in the thermal state, implies that in the state obtained by acting with a right-to-left Wilson line in the charge $n$ representation on the thermofield double, the expectation values of $Q_R,Q_L$ are both given by $n$. This unfortunately does not quite establish that there must be states of charge $n$ for any $n$ however, since thermal charge fluctuations could be large.}  The form of the weak gravity conjecture stated above is essentially a strengthening of the principle of completeness; it also seems to be true in string theory, but again there are not any airtight arguments for its validity.  In fact there is even some debate over what the statement of the conjecture should actually be.  In the original paper it was suggested that merely having a particle whose charge to mass ratio obeys the bound is sufficient, even if it is not of fundamental charge \cite{ArkaniHamed:2006dz}.  This version would not even imply the principle of completeness.  I view the discussion of this paper as thus supporting the stronger version where there must be a fundamentally-charged particle obeying the bound.\footnote{Very recently this conclusion has also been reached via considering the stability of the conjecture to compactification \cite{Heidenreich:2015nta}.  That work also debugs a proposed counterexample from heterotic strings to the stronger weak gravity conjecture advocated here.}  

The second point is interesting in the context of the ``effective conformal field theory'' of \cite{Heemskerk:2009pn,Fitzpatrick:2010zm,Fitzpatrick:2013twa}, which argues that high-dimension operators in the CFT are not needed for describing low-energy physics in the bulk.  Near the vacuum this essentially follows from the usual Wilsonian decoupling of high-energy degrees of freedom.  We see however that in a nontrivial background such as AdS-Schwarzschild, this is no longer true.  On balance this is perhaps a good thing; if AdS/CFT really solves the black hole information problem, as most people expect it does, then the solution should ultimately rely on special short-distance properties of the theory in the bulk such as those of string theory.  Indeed I view it as something of an embarrassment that so much recent work on the subject, including most of my own, has not used bulk UV physics at all (see however \cite{Dodelson:2015toa}).  Optimistically we can perhaps view the resolution of the factorization problem via heavy charges as the simplest example of this; more generally I am confident that any solution of the factorization problem will teach us something about high-energy physics in the bulk.  

The remainder of this paper will explain the above statements in more detail.  In section \ref{facsec}, I will review the issue of Hilbert space factorization in gauge theories, sharpening the factorization problem.  In section \ref{linersec}, I will explain in more detail how the operators in figure \ref{cutfig} are related by renormalization group flow.  In section \ref{iruvsec}, I will discuss the short-distance nature of the factorization problem in more detail, and argue that its resolution requires the gauge field to be emergent.  In section \ref{cpnsec}, I will review the basic properties of the lattice $\mathbb{CP}^{N-1}$ model in $d\geq 2$ dimensions, and show that, although the lattice Hilbert space factorizes, at large $N$ it has a stable Coulomb phase that for $d\geq 4$ persists all the way to the IR and does not confine.  In section \ref{wgcsec}, I will explain how the $\mathbb{CP}^{N-1}$ model, understood as a resolution of the factorization problem, automatically obeys the weak gravity conjecture in any dimension.  I will argue that the reasons for this should apply more broadly, and thus give a new motivation for why some version of the conjecture should be true.  Finally in section \ref{gravsec}, I will give some preliminary comments on the gravitational factorization problem; it seems to have important consequences both for the ``superselection sector'' proposal of Marolf and Wall \cite{Marolf:2012xe} and for reconstruction of the interior in AdS wormholes more generally.

\section{Hilbert space structure in gauge theory}\label{facsec}
\bfig
\includegraphics[height=2cm]{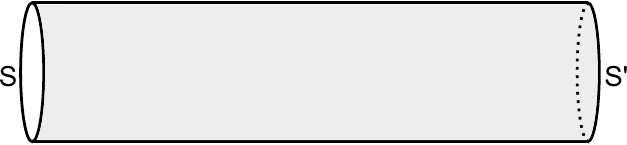}
\caption{Pure electrodyanamics on a spatial cylinder.  The integrated electric flux operator $\int_S E\cdot dA$ through any sphere $S$ is nontrivial, but is independent of which sphere we pick.}\label{cylinderfig}
\efig
Recently there has been a fair bit of interest in the Hilbert space structure of gauge theories \cite{Donnelly:2011hn,Casini:2013rba,Radicevic:2014kqa,Huang:2014pfa,Radicevic:2015sza}, mostly in the context of trying to define the von Neumann entropy of a spatial region.  The standard definition of this entropy would require us to be able to associate a tensor factor of the Hilbert space to the region, but in gauge theories it is not at all clear one can do this.  

As an extreme illustration, consider pure electrodynamics on $\mathbb{R}^2 \times \mathbb{S}^2$, quantized in a Hamiltonian formulation with spatial slices $\mathbb{R}\times \mathbb{S}^2$.\footnote{Details of this formalism and conventions are described in appendix \ref{gaugeapp}}  This situation is shown in figure \ref{cylinderfig}; the topology is the same as that of the four dimensional $AdS$-Schwarzschild solution, which is of course not a coincidence.  To restrict to gauge-invariant states, we need to impose the Gauss constraint
\be\label{gauss}
\nabla \cdot E=0.
\ee
This has the consequence that the integrated electric flux through the sphere is independent of where the sphere is located on the line; this statement holds as an operator equation in the physical Hilbert space!  Moreover this operator is nontrivial, since depending on the choice of boundary conditions we can easily allow for states with nonzero electric flux.  For example we can regulate the spatial line as a finite interval and impose ``perfect conductor'' boundary conditions \eqref{bc} at each end; this is analogous to the standard quantization of a Maxwell field in $AdS$. Such an equivalence between two macroscopically separated operators would be impossible in a theory where we could associate tensor factors to spatial regions, since in a tensor product space the only operator that is shared by both factors is the identity.\footnote{We can describe this nicely using the beautiful algebraic formalism of \cite{Casini:2013rba}, who argue that it is really operator algebras we should associate to spatial regions rather than tensor factors.  We can associate a tensor factor to a region if and only if the algebra we associate to that region has trivial center.  We see in this example that any spatial region of bounded extent which contains a complete sphere will necessarily have a center, since the integrated flux through that sphere is equivalent to an operator far outside the region, and thus lies in the commutant of its algebra.  Unlike in the topologically trivial cases considered in \cite{Casini:2013rba}, it is clear here that no amount of redefining the operator algebra at the boundary of the region will be able to trivialize this center.}  In fact had we assumed that the Hilbert space factorized, we could have argued, as I did in the introduction for AdS/CFT, that charged fields must be present to allow the electric flux at either end of the cylinder to be nonzero.  

\bfig
\includegraphics[height=2cm]{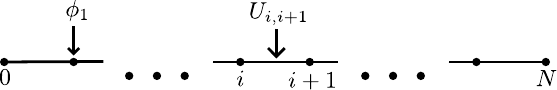}
\caption{Lattice scalar electrodynamics in $1+1$ dimensions.  The gauge field is described by assigning an element of $U(1)$ to each link, while the charged scalar is described by adding a complex number to each site. A gauge transformation $V_i$ assigns an element of the group to each site.}\label{qed2fig}
\efig
The situation becomes more subtle once we include charged fields, since the Gauss constraint \eqref{gauss} then has something nontrivial on the right hand side.  To study this concretely, as in \cite{Harlow:2014yoa} it is very convenient to specialize the discussion to the case of $U(1)$ lattice gauge theory coupled to charged scalars in $1+1$ dimensions.  The basic set of degrees of freedom are illustrated in figure \ref{qed2fig}.  I will take the scalar to be in the fundamental representation, so gauge transformations act as
\begin{align}\nonumber
\phi_i'&=V_i\phi_i\\
U_{i,i+1}'&=V_{i+1}U_{i,i+1}V_i^\dagger.
\end{align}
Our perfect conducting boundary conditions \eqref{bc} allow $U_{0,1}$ and $U_{N-1,N}$ to be arbitrary, but require $V_0=V_N=1$.  For the scalar I'll impose Dirichlet boundary conditions $\phi_0=\phi_N=0$, which again is the natural analogue of standard quantization in $AdS$.  With these boundary conditions, interesting gauge-invariant operators include
\begin{align}\nonumber
W&\equiv U_{0,1}U_{1,2}\ldots U_{N-1,N}\\\nonumber
E_{i,i+1}&\equiv - U_{i,i+1}\frac{\partial}{\partial U_{i,i+1}}\\\nonumber
\overrightarrow{\phi}_i&\equiv \phi_i U_{i,i+1}U_{i+1,i+2}\ldots U_{N-1,N}\\\nonumber
\overleftarrow{\phi}_i&\equiv  U_{0,1}^\dagger U_{1,2}^\dagger\ldots U_{i-1,i}^\dagger \phi_i\\
\rho_i&\equiv \frac{\partial}{\partial \phi_i} \phi_i-\phi_i^\dagger \frac{\partial}{\partial \phi_i^\dagger},
\end{align}
each of which corresponds to one of the gauge invariant operators in the right diagram of figure \ref{giopsfig}.  They are not all independent, for example $\overrightarrow{\phi}_i=W\overleftarrow{\phi}_i$.  

We can write the Gauss constraint as
\be\label{gauss2}
E_{i,i+1}-E_{i-1,i}=\rho_i,
\ee
from which it isn't too hard to show that any gauge-invariant state $|\psi\ran$ in the Hilbert space will have a Schrodinger representation of the form
\be\label{invstate}
\lan U,\phi,\phi^\dagger|\psi\ran=\Psi\left[W,\overrightarrow{\phi_i},\overrightarrow{\phi}_i^\dagger\right].
\ee

Now say that we wish to cut space into two parts, say between site $\ell$ and site $\ell+1$.  If we first consider the case where there is no charged scalar field, then equations \eqref{gauss2}, \eqref{invstate} tell us that the system reduces to a single compact degree of freedom, the Wilson line $W$.  Any attempt to restrict to the observables in some region will eliminate this nonlocal degree of freedom, and we will be left with a trivial Hilbert space.

The situation is not so dramatic when the charged scalar is present, we can define a large algebra of nontrivial gauge-invariant operators that live only on the right side. One set of operators that generates this algebra is $\overrightarrow{\phi}_i$, $\overrightarrow{\pi}_i$, and $E_{i,i+1}$,  with $i>\ell$.  Here $\overrightarrow{\pi}_i\equiv \pi_i U_{i,i+1}^\dagger\ldots U_{N-1,N}^\dagger$, where $\pi_i\equiv -i \frac{\partial}{\partial \phi_i}$.  Similarly we can define an analogous left-sided algebra.  But from \eqref{invstate} we see that we will never be able to generate the Wilson line $W$ out of such operators; this is most clear if we redefine $\Psi$ in \eqref{invstate} to use $\overrightarrow{\phi}_i$ for $i>\ell$ and $\overleftarrow{\phi}_i$ for $i\leq\ell$.  The gauge-invariant Hilbert space thus does not factorize; there is always the remaining degree of freedom $W$.\footnote{We can again demonstrate this very nicely in the language of \cite{Casini:2013rba}; each algebra has a nontrivial central element $E_{\ell,\ell+1}=E_{\ell-1,\ell}+\rho_\ell=E_{\ell+1,\ell+2}-\rho_{\ell+1}$, and thus cannot correspond to a tensor factor.}

Despite this continued non-factorization, I claim that things have improved considerably.  The reason is that, although we cannot produce $W$ out of our left- and right-sided algebras, we \textit{can} produce the operator
\be
W'\equiv \overleftarrow{\phi}_\ell^\dagger\overrightarrow{\phi}_{\ell+1}.
\ee
This operator is not equivalent to $W$, but I claim they are proportional when used in sufficiently low energy states.  In other words, $W'$ flows to $W$ under renormalization group transformations, so their difference is not detectable in low energy experiments. This is not a coincidence: it is somewhat suspicious that, unlike in the charge-free case, we had to go to an explicit lattice description to derive nonfactorization, and in fact in section \ref{cpnsec} we will see that there are regulators of scalar QED where the Hilbert space \textit{does} factorize.  We have thus reduced the factorization problem to a short-distance problem.  Showing the low-energy proportionality of $W$ and $W'$ directly in the lattice theory would be rather tedious, so I will instead now give a continuum argument.

\section{Wilson line renormalization group}\label{linersec}
It is well-known that different line operators can mix under renormalization group flow, see for example \cite{Polchinski:2011im} for a recent discussion and further references.  For our purposes we would like to argue that in electrodynamics, we can make the replacement suggested in figure \ref{cutfig} without affecting low energy correlation functions.  We can formulate this statement as a ``gauge-covariant operator product expansion''
\be\label{ope}
\phi(x)^\dagger \phi(y)=e^{i \int_\Gamma A}\left(G(x,y)+\mathrm{less\,\, singular\,\, terms}\right).
\ee
Here $\Gamma$ is a straight line from $x$ to $y$, and $G(x,y)$ is a $c$-number function that at zeroth order in $q$ is just the free field propagator for $\phi$.  Less singular means less singular in the limit that $x$ and $y$ approach each other.  I will always take $|x-y|$ to be small compared to the inverse mass of $\phi$, so assuming canonical normalization we have
\be\label{scalarprop}
G(x,y)=\frac{1}{2\Delta \Omega_{d-1}}\frac{1}{|x-y|^{2\Delta}}+O(q^2).
\ee  
Here $\Delta\equiv \frac{d-2}{2}$ is the free scaling dimension of $\phi$ and $\Omega_{d-1}=\frac{2\pi^{d/2}}{\Gamma(d/2)}$ is the volume of a unit $\mathbb{S}^{d-1}$.  \eqref{ope} should also be true for other charged operators, with appropriate modifications of $G$ and the representation of the Wilson line.

Equation \eqref{ope} should hold on roughly the same grounds that we expect the ordinary operator product expansion to hold, at least as an asymptotic series \cite{Weinberg:1996kr}.  In conformally invariant gauge theories, it can be derived by moving the Wilson line to the other side of equation to define the gauge-invariant string operator $\phi(x)^\dagger e^{-i \int_\Gamma A}\phi(y)$.  We can then surround this operator with a sphere to define a state in the Hilbert space of the theory on the sphere, which we can then expand in the eigenstates of the dilation operator. Via the state-operator correspondence, this is equivalent to expanding the string operator in a basis of gauge-invariant local operators.  The first term in \eqref{ope} corresponds to the contribution of the identity operator, which should dominate the expansion at small separation.\footnote{The dominance of the identity is certainly true at weak coupling, showing it in general is more subtle.  The ordinary operator product expansion is highly constrained by the assumption that the two operators being multiplied are conformal primaries, which ensures that the identity always dominates at short distance if it appears with nonzero coefficient.  The classification of conformal transformations of finite line operators is more complicated, so far not too much seems to have been said about it (although see \cite{Kapustin:2005py} for some remarks on the infinite case).  What we'd like to say is that there is a basis for ``straight line'' operators, each of which transforms into a rescaled version of itself under a dilation.  We can then use cluster decomposition to argue for short-distance identity dominance for these operators, and thus for any line operator that is a finite superposition of them.  Alternatively, Juan Maldacena has suggesting arguing for the replacement of figure \ref{cutfig} directly by studying the Hilbert space of the CFT on a sphere with two external charges at opposite points, and using the fact that the ground state has the two charges connected by a straight Wilson line.}  This presentation of \eqref{ope} might be called a ``string operator expansion'', in general it is probably the nicer statement, but \eqref{ope} is what we need to address the factorization problem. 

We can also confirm \eqref{ope} at weak coupling by inserting both sides into correlation functions and computing Feynman diagrams.  In the remainder of this section I will illustrate this for scalar QED in $d$ dimensions to leading nontrivial order in the coupling $q$.  In the presence of a Wilson line along a curve $x^\mu(s)$, the Feynman rules are modified by allowing photon propagators to be attached to the line as
\be\label{lineinsert}
i\int ds \dot{x}^\mu(s) G_{\mu\nu}\left(x(s),\cdot\right).
\ee
For example in the free Maxwell theory the vacuum expectation value of a closed Wilson line exponentiates to\footnote{Remember here that in our conventions \eqref{action}, the photon propagator $G_{\mu\nu}$ has an overall factor of $q^2$.}
\be
\lan e^{i\int ds \dot{x}^\mu A_\mu}\ran=e^{-\int ds ds' \dot{x}^\mu(s) \dot{x}^\nu (s')G_{\mu\nu}\left(x(s),x(s')\right)}.
\ee

To check \eqref{ope}, we can compare \eqref{lineinsert} evaluated on the line $\Gamma$ running from $x$ to $y$ to what we get at first order in $q$ from inserting $\phi^\dagger(x)\phi(y)$ and bringing down a single interaction vertex:
\begin{align}\nonumber
iG(x,y)\int_0^1 ds\dot{x}^\mu(s) G_{\mu\nu}\left(x(s),\cdot\right)\stackrel{?}{\approx}i\int d^d z \Big[&G(x,z)\frac{\partial}{\partial z^\mu}G(z,y)-G(y,z)\frac{\partial}{\partial z^\mu}G(z,x)\Big]\\&\times G^{\mu\nu}(z,\cdot).\label{showthis}
\end{align}
To leading order in the separation $|x-y|$, which I will take to be in the $x^1$ direction, we can approximate the external photon propagator on both sides as $G_{\mu\nu}(x,\cdot)$, so it suffices to show that
\be
\frac{1}{G(x,y)|x-y|}\int d^d z \left(G(x,z)\frac{\partial}{\partial z^\mu}G(z,y)-G(y,z)\frac{\partial}{\partial z^\mu}G(z,x)\right)=\delta^1_\mu.
\ee  
Rotational invariance ensures that the $\mu\neq 1$ components will vanish.  Using \eqref{scalarprop}, the $\mu=1$ component boils down to showing that\footnote{In $d=2$ the integral is divergent in the IR; this is a reflection of the standard fact that a massless scalar field does not really exist in $1+1$ dimensions \cite{Coleman:1973ci}.  We could regulate this by reintroducing the mass, but it is easier to instead differentiate both sides of \eqref{ope} with respect to $x$ and $y$; this will then produce an IR-convergent integral and the analogue of \eqref{showthis} will hold automatically by analytic continuation from higher $d$.}
\be
\int_0^\infty d \rho\int_{-\infty}^\infty dh \frac{\rho^{d-2} }{(\rho^2+h^2)^{\Delta}(\rho^2+(h-1)^2)^\Delta}\left(\frac{h}{h^2+\rho^2}+\frac{1-h}{\rho^2+(h-1)^2}\right)=\frac{\Omega_{d-1}}{\Omega_{d-2}},
\ee
which is in fact true.\footnote{One way to evaluate this integral is to observe that for $d>3$, we can consider the two terms in the sum separately. Moreover they are in fact equal, via the change of variables $h\to 1-h$.  We can then use the identity $\frac{1}{x^A}=\frac{1}{\Gamma(A)}\int_0^\infty dt\, t^{A-1}e^{-t x}$ twice to rewrite each factor in the denominator, perform the Gaussian integrals over $\rho$ and $h$, change variables to $a=\frac{tt'}{t+t'}$ and $b=\frac{t}{t+t'}$, and finally perform the now trivial $a$ and $b$ integrals.  This answer then holds for $d>2$ by analytic continuation.}

It would be interesting to study the gauge-covariant operator product expansion \eqref{ope}, or its cousin the string operator expansion, in more detail; I leave this to future work.

\section{UV and IR physics in AdS/CFT}\label{iruvsec}
\bfig
\includegraphics[height=5cm]{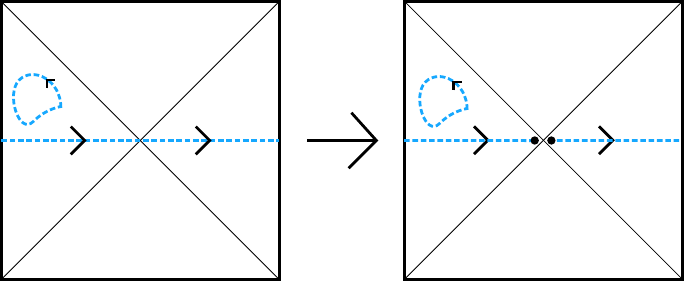}
\caption{Using the replacement of figure \ref{cutfig} to reconstruct a wormhole-threading Wilson line; in the right diagram all operators can be reconstructed using standard techniques.}\label{replacefig}
\efig
Let's now stop and take stock of the situation.  We've argued using the gauge-covariant operator product expansion that by making the replacement of figure \ref{cutfig}, we can split any Wilson line without this being detected in low-energy correlation functions.  This includes the wormhole-threading Wilson line of figure \ref{giopsfig}.  But once we have split it in this manner, we can now reconstruct it in the CFT using the machinery of appendix \ref{gaugeapp}.  This procedure is illustrated in figure \ref{replacefig}.  

This already leads to some interesting conclusions.  First of all clearly a charged field must exist in the bulk if we are to split the Wilson line this way.  We already saw the necessity of this in the introduction on quite general grounds, but now we see something more; the charged field must transform in the fundamental representation of $U(1)$ if we are going to be able cut a fundamental Wilson line.  We thus see that the principle of completeness of \cite{Polchinski:2003bq,Banks:2010zn} comes along for free; the entire charge lattice of the theory will be occupied.\footnote{For example to get the higher charges we can always collapse the fundamental-charge particles into black holes.  From the CFT point of view, we can take the OPE of the charged operator dual to the fundamental field with itself, and with other operators, to generate operators of arbitrary charge.}

Moreover, we see that this construction works even if the charges are quite heavy; we just need to bring them closer together than their inverse mass.  From the bulk point of view this may be surprising; to reconstruct a rather low-energy correlation function involving just Wilson lines and electric fields, we need to know about the existence of very heavy charges.  This is not an \textit{observable} violation of bulk effective field theory, since a bulk effective field theorist can just use the Wilson line directly, but were she to want to use the CFT to compute what she will see, she would have no choice but to use rather high-dimension operators, even though the fields they correspond to have masses that are much larger than what she can test in her laboratory.  This is in some tension with the philosophy of ``effective conformal field theory'' advocated in \cite{Heemskerk:2009pn,Fitzpatrick:2010zm,Fitzpatrick:2013twa}, but it is not a contradiction with the validity of ECFT near the vacuum since we are considering rather high-energy states in the CFTs.  The tension arises because this high energy goes into created a weakly curved background (away from the black hole singularities), rather than any localized high-energy scattering processes, so a bulk observer should still expect to be able to use low-energy effective field theory.  This is another manifestation of the short-distance nature of the factorization problem.  

So have we solved the factorization problem?  No!  We have only postponed it.  What we have achieved is the following: you give me a correlation function containing a wormhole-threading Wilson line that you are interested in, then I can give you an operator which behaves like that Wilson line in that correlation function.  But now say you change your mind, and want to consider a different correlation function.  You can very easily add a few operators near the center of my ``Wilson line'' that will reveal that I am lying!  Of course I can then adjust my ``Wilson line'' to have the charges be close enough to still behave like the actual Wilson line in the presence of your new operators, but now you can just add a few more and catch me again.  To really solve the factorization problem, I must give you a single CFT operator that reproduces the Wilson line \textit{in any situation where it makes sense in the bulk}.  If I am to do this using the mechanism of figures \ref{cutfig},\ref{replacefig}, then naively I need to bring the charges together to within a Planck length to preclude the cat and mouse game just described.  This pushes the problem into the regime of poorly understood bulk physics.  This may be sufficient to count as a solution, but I find it preferable when possible to see if there can be a solution based on well-understood physics.  What I will thus do instead is consider the possibility that the effective Maxwell description of the physics breaks down and some lower energy scale, in a way that can still be described using quantum field theory.  Said differently, the gauge field should be emergent. As long as I bring the charges to within the distance set by that scale, then I no longer can be required to reproduce the physics of the Wilson line any further.\footnote{Of course the first scenario can also be called an emergent gauge field, but it emerges at the Planck scale in some way that also presumably involves the emergence of gravity.  The goal here is, for simplicity, to separate these two phenomenon.  Perturbative string theory seems to be a nice compromise where gauge fields and gravity emerge together, but in a more controlled way at an energy scale below the Planck scale.  It would be very interesting to see if one could understand the factorization problem more precisely in that language; I'll discuss this a bit more in section \ref{wgcsec}.}

To sharpen this discussion, I will now discuss in detail a field theory where we indeed have an emergent gauge field at long distances but a factorizable Hilbert space at short distances.  We will see that it gives a concrete realization of the philosophy of the previous paragraph; I view it as an existence proof for a resolution of the factorization problem based on the idea of figures \ref{cutfig},\ref{replacefig}.

\section{Emergent gauge fields and factorization in the \cpn model}\label{cpnsec}
The \cpn model has a long history. In $1+1$ dimensions its large $N$ solution was first described in \cite{D'Adda:1978uc,Witten:1978bc}.  In $2+1$ dimensions it has been studied extensively by condensed matter physicists in the context of quantum phase transitions in valence bond solids, see for example \cite{senthil2004deconfined,senthil2004quantum,Metlitski:2008dw,Dyer:2015zha} and the references therein.  Euclidean lattice versions of the \cpn model were studied in \cite{Stone:1978pe,Rabinovici:1980dn}; these are equivalent to the version I describe here up to irrelevant operators.  My discussion basically follows \cite{Polyakov:1987ez}; there only the $1+1$ case is discussed, but methods are presented for other models that allow easy generalization to any spacetime dimension.  The only real novelties here are the phase diagrams in figure \ref{phasesfig} and the discussion of the emergence of the Wilson line and the resolution of the factorization problem in the last subsection. 

The reader should keep in mind that I am always viewing the \cpn theory as a model of the \textit{bulk} physics of the factorization problem, not of the boundary CFT.

\subsection{The model}
The \cpn model is a quantum field theory of $N$ complex scalar fields $z_a$, obeying a constraint $\sum_{a=1}^N z^*_a z_a=1$.  It has a global $SU(N)$ symmetry under which $z_a$ transforms as a fundamental.  Moreover we impose a gauge symmetry under 
\be
z'_a(x)=e^{i\theta(x)}z_a(x),
\ee
which, together with the constraint, means that the theory really describes a nonlinear-$\sigma$ model with target space \cpn.  From now on I will use matrix notation to suppress the $SU(N)$ index, for example I'll write the constraint as $z^\dagger z=1$.
The Lagrangian in the continuum is
\be
\mathcal{L}=-\frac{N}{g^2}\left(D^\mu z\right)^\dagger D_\mu z,
\ee
where the covariant derivative $D_\mu \equiv \partial_\mu-i A_\mu$ is defined using 
\be\label{Aeq}
A_\mu \equiv \frac{1}{2i}\left(z^\dagger \partial_\mu z-\partial_\mu z^\dagger z\right).
\ee
In the Hamiltonian formalism the momentum conjugate to $z$ is
\be
\pi=\frac{N}{g^2}\left(D_0 z\right)^\dagger,
\ee
and the Hamiltonian density is
\be
\mathcal{H}=\frac{g^2}{N}\pi\pi^\dagger+\frac{N}{g^2}\left(D_i z\right)^\dagger D_i z+iA_0\left(\pi z-z^\dagger \pi^\dagger\right).
\ee
The term multiplying $A_0$ is an additional constraint, analogous to the Gauss constraint \eqref{gauss}.  In the Hamiltonian quantization of this theory we have the primary constraint $z^\dagger z=1$, the secondary constraint $\pi z+z^\dagger \pi^\dagger=0$, and this Gauss constraint.  We can implement the first two by doing Dirac quantization, and then use the third to project onto physical states.  Alternatively we could introduce a local gauge-fixing condition and then treat all four simultaneously using Dirac quantization.  Either way, the key point is that all the constraints are ultralocal.  This means that a lattice regularization of this theory will have a Hilbert space that automatically factorizes; an orthonormal basis is given by specifying an element of \cpn at each point on the spatial lattice.  I'll defer the details of this lattice theory to section \ref{cpfactorsec} below.

\subsection{The solution}
At large $N$ the \cpn model can be solved by a standard set of tricks \cite{D'Adda:1978uc,Witten:1978bc,Polyakov:1987ez}.  The basic idea is to introduce an auxiliary gauge field $A_\mu$, whose equation of motion is given by \eqref{Aeq}, and a Lagrange multiplier $\sigma$ to impose the constraint $z^\dagger z=1$.  The Lagrangian becomes
\be
\mathcal{L}=-\frac{N}{g^2}\left((D^\mu z)^\dagger D_\mu z+\sigma(z^\dagger z-1)\right),
\ee
where now $z$ is unconstrained and $A_\mu$ in the covariant derivative is taken as an independent field. It is convenient to look at the Euclidean path integral formulation of this theory, where we want to compute\footnote{Deriving this path integral from the Hamiltonian quantization is nontrivial, because of the constraints.  We should in principle worry about additional terms in the action being generated by the functional determinants that arise \cite{Senjanovic:1976br}, but fortunately they are all field-independent and can be ignored.  Starting from the path integral point of view, the naive definition of the measure is already invariant under the global $SU(N)$ symmetry and local $U(1)$ symmetry, and we have already included all relevant and marginal terms allowed by these symmetries.}
\be
Z=\int \mathcal{D}A \mathcal{D}\sigma \mathcal{D}z \mathcal{D}z^\dagger e^{-\frac{N}{g^2}\int d^d x\left((D^\mu z)^\dagger D_\mu z+\sigma(z^\dagger z-1)\right)}.
\ee
At fixed $A_\mu$ and $\sigma$, the integrals over $z$,$z^\dagger$ are Gaussian and can be evaluated to give an effective action for $A_\mu$ and $\sigma$.  At large $N$ this effective action will be exact, since the fluctuations of $A_\mu$ and $\sigma$ will only shift it by terms that are subleading in $1/N$.  Formally this effective action is given by
\be\label{Seff}
S_{eff}[A,\sigma]=N\left[-\frac{1}{g^2}\int d^d x \sigma+\log \det \left(\frac{-D^2+\sigma}{\Lambda^2}\right)\right],
\ee
where $\Lambda$ is the UV cutoff, set by the inverse lattice spacing.  

To understand the dynamics implied by \eqref{Seff}, the first order of business is to study the effective potential for $\sigma$, obtained by setting $A_\mu\to 0$ and taking $\sigma$ to be constant.  This is extremized at some $\sigma_0$, determined by the ``gap equation''
\be\label{gap}
\frac{1}{g^2}=\int \frac{d^d p}{(2\pi)^d}\frac{1}{p^2+\sigma_0}.
\ee
The integral on the right hand side is UV divergent for $d\geq 2$, so we will always cut it off at $p^2=\Lambda^2$.  For $d=2$ this equation then has a nonzero solution for any real value of $g$, given by 
\be
\sigma_0=\Lambda^2 e^{-\frac{4\pi}{g^2}}.
\ee
This nonzero expectation value for $\sigma$ gives a mass to the $z$ particles, which is exponentially small via dimensional transmutation.  

For $d>2$, the right-hand side of the gap equation \eqref{gap} is bounded from above by some finite value, so for sufficiently small $g$ there is no real solution for $\sigma_0$.  This reflects the basic observation that at sufficiently small $g$, the global $SU(N)$ symmetry is spontaneously broken, as we would expect for a weakly coupled nonlinear-$\sigma$ model in $d>2$.  So rather than interpreting $\sigma_0$ as a mass term, we need to re-expand around the true vacuum at some $z\neq 0$.  This also Higgses the $U(1)$ gauge symmetry.  At a particular critical value of $g$, given by
\be
\frac{1}{g_c^2}=\int \frac{d^d p}{(2\pi)^d}\frac{1}{p^2},
\ee
we have $\sigma_0=0$ and the system becomes scale invariant.  For $g>g_c$ we have $\sigma_0>0$ so the $SU(N)$ symmetry is unbroken.  For $d=2$ we have $g_c=0$, as required by the Mermin-Wagner-Coleman theorem \cite{Mermin:1966fe,Coleman:1973ci}.

\bfig
\includegraphics[height=2cm]{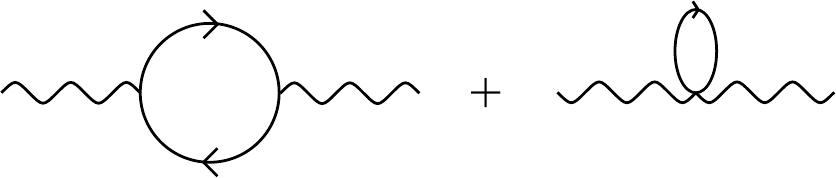}
\caption{Diagrams for the quadratic part of the gauge field effective action at large $N$.}\label{diagramsfig}
\efig
The ordered phase with $g<g_c$ is uninteresting for our purposes, since the gauge field is Higgsed and the low energy physics is described by weakly interacting Goldstone bosons.  For $g>g_c$ we can potentially generate interesting dynamics.  By bringing $g$ close to $g_c$ from above, we can arrange for $\sigma_0$ to be much less than $\Lambda^2$, in which case we can study the system using continuum methods.  In particular we can compute the part of $S_{eff}$ that is quadratic in $A_\mu$; this is given by
\begin{align}\nonumber
S_{eff}\supset-N\int \frac{d^d\ell}{(2\pi)^d} A_\mu(\ell)A_\nu(-\ell)\int \frac{d^d p}{(2\pi)^d}\Bigg[&\frac{(2p^\mu+\ell^\mu)(2p^\nu+\ell^\nu)}{(p^2+\sigma_0)((p+\ell)^2+\sigma_0)}\\
&-2\delta^{\mu\nu}\frac{1}{p^2+\sigma_0}\Bigg].
\end{align}
The right-hand side of this equation comes from summing the Feynman diagrams shown in figure \ref{diagramsfig}.  The integral over $p$ can be evaluated using the usual Feynman parameter technology.  We are especially interested in the low-energy piece proportional to $\ell^\mu \ell^\nu-\delta^{\mu\nu} \ell^2$, since this gives rise to an effective Maxwell term.  Computing the coefficient of this term, we find the low-energy gauge couplings
\be\label{couplings}
\frac{1}{q^2}=
\begin{cases}
\frac{N}{6\pi \sigma_0} & d=2\\ 
\frac{N}{12\pi \sqrt{\sigma_0}} & d=3\\
\frac{N}{12\pi^2}\log \left(\frac{\Lambda}{\sqrt{\sigma_0}}\right) & d=4 \\
N\Lambda^{d-4} & d>4.
\end{cases}
\ee
For $d\leq 4$ these can be computed using dimensional regularization. For $d>4$ the integral has a power-law UV divergence, so the overall normalization is non-universal; I've simply fixed the power of $\Lambda$ by dimensional analysis.  For $d\geq 4$ we cannot take the continuum limit and still obtain a finite coupling; this reflects the usual short-distance misbehavior of electrodynamics.
    
The main lesson of \eqref{couplings} is of course that $\frac{1}{q^2}$ is finite in all cases; we have generated Maxwell dynamics for $A_\mu$, which has thus become a real gauge field mediating Coulomb interactions between the massive charged $z$'s!\footnote{One way to see that the $z$'s in fact correspond to massive particles is to not integrate over a few of them.  The integral over the rest of them will not generate any new terms in the effective action for the ones we kept around since they are decoupled.  One can also check that $\sigma$ does not lead to any new particles in the infrared.}  
Higher derivative corrections are suppressed by the scale $\sqrt{\sigma_0}$, so the Maxwell term does not dominate the behavior of $A_\mu$ for energies above this.  

For $d=2,3$, these Coulomb interactions eventually become strong and confine the charged $z$'s into neutral objects.  In $d=2$ this happens classically, since the Coulomb potential is linear.  In $d=3$, monopole operators eventually proliferate in the path integral over $A_\mu$, again leading to a linear potential \cite{Polyakov:1987ez}.  For $d\geq 4$, whether the theory confines depends on how large of a coupling is generated; there is a phase transition at finite $q$.

\bfig
\includegraphics[height=5cm]{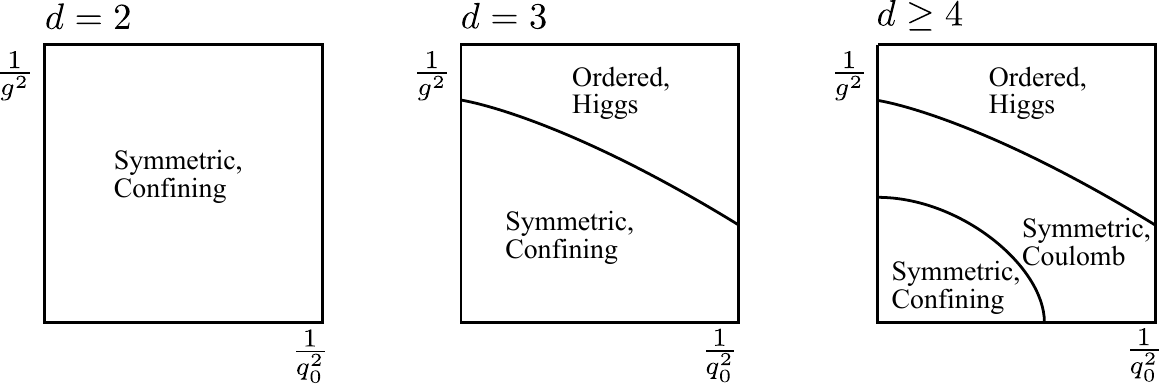}
\caption{Phase diagrams for the large-$N$ lattice \cpn-Maxwell system in various dimensions.  The left side is the pure \cpn model discussed in the text, the bottom is pure $U(1)$ lattice gauge theory, and the right side is the $\mathbb{S}^{2N-1}$ nonlinear-$\sigma$ model.  The phases are labeled by first saying what happens to the $SU(N)$ global symmetry, and then the $U(1)$ gauge field.  In the text we are predominantly interested in approaching the (upper) critical point on the left side from below.}\label{phasesfig}
\efig
We can summarize all of the results of this subsection into a phase diagram for the large $N$ lattice \cpn model, shown for various dimensions in figure \ref{phasesfig}.  To make the physics more clear, I have included a bare Maxwell term $-\frac{1}{4q_0^2}\int F^2$ (or more precisely its lattice analogue, the sum over all single-plaquette Wilson loops \cite{Wilson:1974sk}), and shown the phase diagram as a function of both $\frac{1}{g^2}$ and $\frac{1}{q_0^2}$. This phase diagram is reminiscent of that studied in \cite{Fradkin:1978dv,Banks:1979fi} for lattice gauge theory with Higgs fields.\footnote{I thank Steve Shenker for suggesting this comparison, which proved extremely useful in organizing my thinking on the lattice \cpn model for $d>2$.}   It is interesting to note the difference of the $d\geq 4$ diagram from that constructed in \cite{Fradkin:1978dv,Banks:1979fi} for $U(1)$ gauge theory with a fundamental Higgs field; namely for the \cpn model the Coulomb phase persists all the way over to $q_0=\infty$, while in \cite{Fradkin:1978dv,Banks:1979fi} the Higgs and confining phases were continuously connected.  This is not possible here, since the global $SU(N)$ symmetry is broken in the Higgs phase and restored in the confining phase.

\subsection{The factorization problem}\label{cpfactorsec}
We have now seen that the lattice \cpn model has a nontrivial Coulomb phase at large $N$, which is stable for $d\geq 4$ and which undergoes Abelian confinement for $d=2,3$.  It thus has nontrivial Wilson line operators in the infrared whose existence must be consistent with the microscopic tensor factorization of the Hilbert space site by site.  It is quite illuminating to see how these Wilson lines can be written in terms of the lattice variables.  This amounts to looking for a lattice version of eq. \eqref{Aeq}.  

I first need to establish a bit of notation: I will denote site vectors on the lattice as $x$, single-site displacement vectors as $\delta$, and links as $x$,$\delta$.  To avoid overcounting of links I will restrict $\delta$ to point in a direction of coordinate increase.  Now say I have two neighboring sites $x$ and $x+\delta$.  We can then define the link variable
\be\label{linkcut}
U_{x,\delta}\equiv\frac{z^\dagger_{x}z_{x+\delta}}{|z^\dagger_{x}z_{x+\delta}|},
\ee
which indeed reduces to \eqref{Aeq} in the limit that $z_{x+\delta}\approx z_x$.  I'll enjoy this equation more in a moment, but to first finish defining the lattice theory the discrete covariant derivative along the link $x,\delta$ can now be written as 
\be
D_{\delta}z_{x}\equiv z_{x+\delta}U_{x,\delta}^*-z_{x}.
\ee
Up to a constant shift, and setting the Gauss constraint to zero, the lattice Hamiltonian is then
\be \label{latticeH}
H=\frac{g^2}{N}\sum_{x}\pi_x \pi^\dagger_x+\frac{N}{g^2}\sum_{x,\delta}\left(z^\dagger_{x+\delta}U_{x,\delta}z_x+c.c.\right).
\ee
This is the real starting point for the phase diagram \ref{phasesfig}.\footnote{As in the continuum version of the model, we can also instead take $U$ to be an independent variable and integrate over it.  This makes the derivation of the phase diagram easier, since in the $g\to \infty$ limit we directly recover the standard Abelian lattice gauge theory and can use known results.  Performing this integral first reproduces \eqref{latticeH} up to finite renormalizations and irrelevant operators, so the phase diagram is the same as in the theory with only the $z$'s.}

Returning now to eq. \eqref{linkcut}, we see that the Wilson link operator is explicitly defined as a contraction of charged fields: the \cpn model thus builds the Wilson line out of the cutting operation of figure \ref{cutfig} from the beginning!  Said differently, eq. \eqref{linkcut} shows that the expansion \eqref{ope} becomes exact at the lattice scale.  The only new feature is that, in order to obtain an $SU(N)$ singlet, we need to sum over flavors.  An experimentalist who attempted to probe this Wilson line at energy scales above $\sqrt{\sigma_0}$ would begin to see deviations from Maxwell behavior, and eventually it would completely disassociate at the lattice scale.  We thus see that indeed the \cpn model explicitly implements the philosophy of the previous section.

\section{The weak gravity conjecture}\label{wgcsec}
We've now seen that the factorization problem can be resolved entirely within effective field theory; we are able to have a microscopic Hilbert space that factorizes into regions but nonetheless have nontrivial Wilson line dynamics in the infrared.  This resolution requires the existence of charged particles in the fundamental representation, and thus implies the completeness principle of \cite{Polchinski:2003bq,Banks:2010zn}.  It is natural to wonder if it also has something to say about a well-known strengthening of this principle, the ``weak gravity conjecture'' of \cite{ArkaniHamed:2006dz}.   

Roughly speaking, the weak gravity conjecture says that, in any theory whose low energy limit includes Einstein gravity coupled to a Maxwell field, if the gauge coupling is weak then there must be a charged particle whose mass is small in Planck units.  The original argument given for this is that otherwise no extremal charged black hole would be able to decay, which would lead to a weird situation where there are large numbers of stable states that are not protected by any symmetry (those which are protected by supersymmetry would be stable regardless).  Of course sometimes life is weird, which is why it remains a conjecture.\footnote{One can try to better justify the conjecture by relating it to well-known arguments against remnants and continuous global symmetries \cite{Banks:2006mm}, but new subtleties arise.  For example one can attempt to produce an object that exceeds the Bekenstein-Hawking entropy, but unlike in a remnant scenario, where unitary evaporation requires the entropy of the remnant to be infinite, here it can exceed the Bekenstein-Hawking entropy only by a finite amount, only in a regime ($m\sim m_p$) where it is not clear we should worry, and only if we define the ensemble the entropy is counting to be rather broad in charge.}  
%Indeed it should be emphasized that the weak gravity conjecture is on less firm footing than other well-known conjectures based on superficially similar arguments.  For example the stable extremal black holes its negation would require are sometimes called remnants, in hopes of using the well-known problems with remnants \cite{Susskind:1995da} to put the conjecture on firmer footing. A remnant however is an object whose entropy exceeds that of Bekenstein and Hawking; these are not, and it is not at all clear that they should lead to similar problems.\footnote{One can try to argue that the Bekenstein-Hawking entropy can be exceeded for black holes whose mass is sufficiently close to $m_p$, but  }  It is also sometimes mistakenly said that this argument for the weak gravity conjecture is comparable to the black hole argument that there are no continuous global symmetries; this too is false.  Assuming that black hole evaporation preserves a continuous global symmetry leads to an actual contradiction with the no-hair theorems of general relativity; the negation of the weak gravity conjecture implies no such contradiction as far as we know.  
In fact its originators were not even sure themselves what the conjecture should say in detail; the weakest version would simply require a particle whose charge to mass ratio exceeds $\sqrt{G}$, while we could also demand it to be the lightest charge particle, or a particle of fundamental charge.  They rejected the last option based on a claimed counterexample from the heterotic string, but this counterexample was recently debugged in \cite{Heidenreich:2015nta}, who also argued that, although the ``no stable  non-supersymmetric extremal black holes argument'' naively only requires the charge to mass ratio to obey \eqref{wgc}, upon dimensional reduction it actually requires the stronger version of \eqref{wgc} where the particle must have fundamental charge.  This will be the version which, for other reasons, I will focus on here; more precisely it says that there must be a fundamentally-charged particle of mass $m$ that obeys\footnote{There is a convention-dependent order-one factor here I have ignored, which is the same as that appearing in the charge to mass ratio of an extremal black hole.  I will not try to reproduce it, so I will not bother to keep track of it.}
\be\label{wgc}
m\leq \frac{q}{\sqrt{G}}, 
\ee
where $G$ is Newton's constant and $q$ is again the gauge coupling.  

Taken together, \cite{ArkaniHamed:2006dz} and \cite{Heidenreich:2015nta} show that the conjecture holds in all known compactifications of string theory that lead to $U(1)$ gauge fields, which by itself is perhaps already evidence enough to take it seriously.  The goal of the rest of this section will be to argue that the factorization problem gives another reason.\footnote{One could also try to prove the conjecture directly in AdS/CFT by using the conformal bootstrap to show that the existence of a current requires charges whose scaling dimensions obey an appropriate version of \eqref{wgc}.  As already mentioned, even the existence of the charges has not been established outside of $1+1$ dimensions, and the validity of \eqref{wgc} seems like it will require special properties of holographic theories \cite{Nakayama:2015hga}.}  

We saw in the previous section that, in a theory with an emergent gauge field, $q$ and $m$ are both computable in terms of the short-distance parameters of the theory.  In any such model we can thus check whether or not \eqref{wgc} holds directly, provided that we can actually compute them. I will now do this for the \cpn model.  Before doing so, I need to rewrite \eqref{wgc} in terms of the \cpn variables.  At first it seems that, since the \cpn model is just a quantum field theory, there is nowhere for Newton's constant $G$ to come from.  We can weakly couple the model to gravity, but we can just take the Planck mass to be arbitrarily high compared to any scale in the model, which would allow \eqref{wgc} to be satisfied trivially.  The conjecture will be tested most strenuously if we take the cutoff $\Lambda$ of the \cpn model to be of order the Planck scale, so that is what I will do (we can't take it to be higher than the Planck scale, since then we would not have solved the factorization problem!).  There is some subtlety in doing this however, since the $N$ charged fields of the model will also renormalize the Einstein-Hilbert term via the diagrams of figure \ref{diagramsfig}.  For $d>2$ they will generate an effective Newton constant of order
\be
\frac{1}{G}\sim N\Lambda^{d-2},
\ee
where the more singular high-energy behavior compared to \eqref{couplings} is of course due to the extra powers of $p$ in the gravitational couplings compared to those of electromagnetism.  This is the well-known lowering of the strong-coupling scale of gravity in the presence of a large number of species \cite{Susskind:1995da,Dimopoulos:2005ac,Calmet:2008tn}.  We thus want to show that
\be
m\leq q \sqrt{N}\Lambda^{\frac{d-2}{2}}.
\ee
Using \eqref{couplings}, it is easy to see that this is indeed true for $d\geq 2$.  We thus see that the electromagnetism generated by the \cpn model automatically obeys the weak gravity conjecture!  

Let's now step back and understand more generally why this happened.  The essence of the weak gravity conjecture is the statement that a weak gauge coupling requires charges whose mass is small in Planck units.  What we learned from the factorization problem, at least if its resolution is to be described by effective field theory, is that at some high energy scale, at most of order the Planck scale, the coefficient of the Maxwell term in the effective action is zero.  If we now wish to have weakly coupled gauge fields at long distances, we need to generate a Maxwell term by integrating out massive charged fields.  But if their masses are too large, they will gap out before they are able to generate a sizeable Maxwell term, and the gauge field will stay strongly coupled.  Moreover the charges will need to be in the fundamental representation, since otherwise we will not be able to use them to split a Wilson line in the fundamental representation.

Does this mean we have derived the conjecture \eqref{wgc}?  I would be surprised if there was a quantum field theory resolution of the factorization problem that did not obey the conjecture, basically for the reasons just given.\footnote{It would be interesting to check this in other examples of theories with emergent gauge fields, for example in supersymmetric examples.  One discussion of some of the relevant issues is \cite{Rabinovici:2011jj}.}  But nobody has promised us that the resolution of the factorization problem must be describable in quantum field theory; indeed in perturbative string theory we know that it isn't!  There the gauge fields and gravity instead emerge together from a nonlocal theory.  Nonetheless our main conclusions from the field theory resolution still seem to be true: there are always fundamentally charged objects, and they always seem to obey \eqref{wgc}.  Perhaps there is a way to describe the emergence of the gauge field in string theory that is more similar to the discussion of the previous paragraph, after all in the usual perturbative formulation it is far from clear that string theory by itself \textit{does} solve the factorization problem of AdS/CFT.  This seems like a question that deserves further study.  

Before concluding this section, I will briefly comment on a previous attempt to understand the weak gravity conjecture purely in terms of the consistency of effective field theory \cite{Cheung:2014vva,Cheung:2014ega}.  There it was first argued that the conjecture is in tension with naturalness, since for a small gauge coupling it might require light scalars.  Here we have seen that the \cpn model will produce a small gauge coupling \textit{only} if the scalars are light.  There is no problem with naturalness, the conjecture is obeyed whether or not we tune the scalars to be light.  Secondly they argued that infrared consistency of the effective field theory of gravitons and photons requires the conjecture to hold, at least when a parameter they called $\gamma$ is small.  But it is clear there can be no such argument in general; for example if we ignore the factorization problem, there is nothing to stop us from adding a huge bare Maxwell term to the action of the \cpn model coupled to gravity.  In the infrared this will produce a perfectly consistent theory that violates the conjecture; and indeed one can check that in the language of  \cite{Cheung:2014ega} this makes $\gamma$ large.  

%In fact the parameter $\gamma$ has a universal $O(1)$ contribution from graviton exchange, so getting into the regime $\gamma\ll 1$ in the first place requires some sort of tuning that cancels this contribution with hypothetical Planck-scale physics.  

%In fact it seems that what is meant by the smallness of $\gamma$ in \cite{Cheung:2014ega} is really just the statement that $m\sqrt{G}\ll q$ for the charges that are integrated out, or in other words the conjecture whose consistency is supposedly being tested.

\section{Comments on the gravitational case}\label{gravsec}
I'll conclude this paper with a few comments on the gravitational version of the factorization problem.    Similar to the Gauss constraint \eqref{gauss} in electrodynamics, in general relativity we have the Hamiltonian and shift constraints, which again involve spatial derivatives (see for example \cite{Kabat:2013wga} for a recent discussion in AdS/CFT).  The tensor product structure of the two CFTs dual to the wormhole geometry again suggests that these constraints must somehow be modified at short distances to avoid a contradiction: gravity must also be emergent in a way that is consistent with this factorization. 

One immediate issue is that, once gravitational backreaction is included, it is no longer clear ``where'' we should view the tensor factorization of the CFTs as splitting the bulk geometry.  A natural guess, already pointed out in \cite{Harlow:2014yoa}, is that it happens at the compact extremal surface of minimal area homologous to one of the boundaries.\footnote{If there is only one such surface, it will also be homologous to the other boundary.  In all the examples I know of wormholes with CFT descriptions, in particular those of \cite{Shenker:2013yza}, this is the case.  There are some more general wormholes which do not have this property, but perhaps none of them really exist as states in the two-CFT Hilbert space.  This is analogous to what one might expect for ``bag of gold'' solutions.}  The electric flux through the bifurcation surface in the electromagnetic case then has a beautiful analogue here; the area (or more generally the Noether charge) of the extremal surface.  This is precisely what appears in the leading-order formula for the entanglement entropy of one of the boundaries \cite{Ryu:2006bv,Hubeny:2007xt}; this suggests a possibly deep connection between the Ryu-Takayanagi/Hubeny-Rangamani-Takayanagi formula and the factorization problem.\footnote{It is worth emphasizing that if we should indeed understand the factorization as happening at this extremal surface, in order to reconstruct the Wilson line we still need to solve the problem of ``entanglement wedge'' reconstruction \cite{Czech:2012bh,Wall:2012uf,Headrick:2014cta}.  Based on recent developments \cite{Almheiri:2014lwa,Jafferis:2014lza,Pastawski:2015qua}, I am optimistic this will happen soon.}

Based on the previous point, the gravitational analogue of the wormhole-threading Wilson line of figure \ref{giopsfig} should then be closely related to the ``opening angle'' variable of \cite{Carlip:1993sa}, evaluated at this extremal surface.\footnote{I thank Don Marolf and William Donnelly for suggesting this.}  We've already seen that the existence of the Wilson line is something of a diagnostic for whether the two sides are connected \cite{Harlow:2014yoa,Engelhardt:2015fwa}; so it must be for its gravitational analogue.  Indeed these gravitational degrees of freedom seem to be precisely those which sew the two sides together, so understanding their description in the CFTs is maximally interesting from the point of view of understanding how the geometric connection emerges from the entanglement of the boundary CFTs \cite{VanRaamsdonk:2010pw}.  Based on what happened in the electromagnetic case, it seems that understanding bulk short-distance physics will be essential.

Continuing in this direction, in \cite{Marolf:2012xe} it was argued that, in the full theory of quantum gravity with two asymptotically-AdS boundaries, in addition to the two CFTs there must also be ``superselection sector'' degrees of freedom which tell us whether or not the two boundaries are connected by a wormhole.  In fact in \cite{Harlow:2014yoa} one possible resolution of the factorization problem that was considered was that these additional degrees of freedom break the factorization in just such a way that the Wilson line can exist.  But by now I hope I have at least made it rather plausible that a wormhole-threading Wilson line can be described entirely within the factorized Hilbert space of the two CFTs; this suggests the same will be true for its gravitational analogue.  If so, it then seems that we do not need superselection sectors to tell us whether or not the bridge is connected.  The CFTs force our hand, by producing a UV completion of gravity where the bridge forms automatically.  

\bfig
\includegraphics[height=5cm]{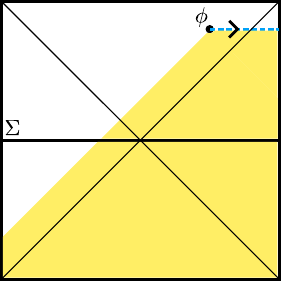}
\caption{Reconstructing an interior operator in the thermofield double state.  We evolve it back to the Cauchy surface $\Sigma$, where it is nontrivial only in the overlap with the yellow region.  All operators on $\Sigma$ must thus be dressed to the right, which requires us to use wormhole-threading Wilson lines.}\label{dressingfig}
\efig
Finally, I would like to emphasize that wormhole-threading Wilson lines, and their gravitational analogues, are very important in defining observables behind the horizon in the neighborhood of the thermofield-double state.  For example, returning to the electromagnetic case for a moment, consider a charged scalar field $\phi$ behind the horizon, shown in figure \ref{dressingfig}.  To make it gauge-invariant, we need to attach some sort of dressing, which we can for simplicity take to be a Wilson line that extends either to the right boundary or to the left boundary.  Either choice may seem a bit counterintuitive from the point of view of the free field limit; one might want to think of the right(left)-moving part of $\phi$ as ``having come from the left(right) boundary'', and thus being naturally dressed by a Wilson line to the left(right), while here I am choosing them both to be dressed in the same direction.\footnote{It is a bit subtle to explicitly describe a dressing where we dress left and right movers in different directions, and in the gravitational case it ends up defining an operator which isn't even associated to a definite point in spacetime, since in the presence of backreaction the two parts may no longer meet at the same point (see \cite{Papadodimas:2015xma} for a similar point).}  The reason for this is that the experiment I am ultimately interested in describing involves an observer who jumps in from one side, say the right side.  All of her experimental equipment will be carried in from the right side, so she must ultimately be interested in operators whose electromagnetic (and gravitational) dressing goes only to that side (see \cite{Susskind:2013lpa,Papadodimas:2015xma} for some similar remarks).  As mentioned in the introduction, the operators which are relevant for observers who do not jump in at very late times can be represented in terms of a complete set of operators on a Cauchy slice $\Sigma$ that passes through the bifurcation surface and does not enter the interior.  In representing $\phi$ however, we should only need to use operators on $\Sigma$ that are nontrivial in its intersection with the past lightcone of $\phi$ and its dressing, shown in yellow in figure \ref{dressingfig}.  We will be interested in charged operators on $\Sigma$ that lie to the left of the bifurcation surface in doing this, but we apparently want to dress them to the \textit{right} to preserve commutativity with operators outside of the yellow region.  But these are precisely the operators we need our wormhole-threading Wilson lines to produce in the CFT; we need to take their left-dressed versions and then use a Wilson line to flip the dressing via $\overrightarrow{\phi}=W\overleftarrow{\phi}$!

Thus we have learned that, at least in the vicinity of the thermofield double state, the CFT descriptions of observables behind the horizon that are relevant for infalling observers \textit{require} the wormhole-threading Wilson line we have studied in this paper if they are charged, and even if they are not they still require its gravitational analogue.  We have thus learned that their description will ultimately rely on understanding short-distance physics in the bulk.  This will continue to be the case in any state where the extremal surface is visible to an infalling observer, while if this is not true then it is unclear whether we should really think of the wormhole as existing at all.  Perhaps further investigation of this issue will shed light on the paradoxes of \cite{Almheiri:2013hfa,Marolf:2013dba,Papadodimas:2015xma}; at a minimum it is probably good news that a deeper knowledge of high-energy physics in the bulk seems to be required to properly discuss these questions.

\paragraph{Acknowledgments} I'd like to thank Nima Arkani-Hamed, Tom Banks, Horacio Casini, Cliff Cheung, Xi Dong, William Donnelly, Thomas Dumitrescu, Ben Freivogel, Steve Giddings, Marina Huerta, Sung-Sik Lee, Hong Liu, Juan Maldacena, Don Marolf, Joe Polchinski, Matt Reece, Grant Remmen, Subir Sachdev, Nati Seiberg, Steve Shenker, Lenny Susskind, Mithat Unsal, Herman Verlinde, Aron Wall, and Sasha Zhiboedov for useful discussions about issues related to this work.  I'd also like to thank the Aspen Center for Physics for hospitality, where a preliminary version of this work was presented in the summer of 2014.  I'd also like to thank the KITP for hospitality, on multiple occasions during the ``Quantum gravity from the UV to the IR'' workshop.  Finally I'd like to thank the theoretical physics community in Princeton for a delightful three years, during which much of this work was carried out.  I am supported by DOE grant  DE-FG0291ER-40654 and the Harvard Center for the Fundamental Laws of Nature, and previously was also supported by the Princeton Center for Theoretical Science.  

\appendix
\section{Notation and review of gauge field reconstruction}\label{gaugeapp}
In this appendix I review electrodynamics in curved spacetime, mostly to establish conventions, and then review the CFT reconstruction of a gauge field in AdS/CFT.
\subsection{Conventions for electrodynamics in curved space}
In this paper I will always normalize any gauge field so that a field of minimal nonzero charge is acted on by the covariant derivative $D_\mu\equiv\nabla_\mu-i A_\mu$.  The strength of electromagnetic interactions will be controlled by the coefficient of the Maxwell term in the action
\be\label{action}
S=-\int d^d x \sqrt{-g}\left(\frac{1}{4q^2}F_{\mu\nu}F^{\mu\nu}- A_\mu J^\mu\right).
\ee
Here $J^\mu$ is the electromagnetic current in the bulk, not to be confused with the CFT current dual to $A_\mu$, which I will denote $j$.  Writing $J^\mu$ out explicitly in terms of canonically-normalized matter fields, it will not have an overall factor of $q$.  The variation of this action leads to Maxwell's equation
\be
\nabla_\nu F^{\mu\nu}=q^2J^\mu,
\ee
together with the requirement that at any spatial boundary we have 
\be
r_\mu F^{\mu\nu}\delta A_\nu=0,
\ee
where $r_\mu$ is the outward pointing normal to the boundary.  In this paper we will always satisfy this by imposing the boundary condition
\be\label{bc}
A_\mu \propto r_\mu,
\ee
since this is the appropriate choice at the boundary of $AdS$ for the standard quantization of the gauge field.  Physically it corresponds to viewing the boundary as a perfect conductor.  This boundary condition is only preserved by gauge transformations that become constant at the boundary, so we must not quotient by gauge transformations that do not.  In fact usually we also do not quotient by gauge transformations that approach a nontrivial constant; these instead become a $U(1)$ global symmetry in the CFT, which acts nontrivially on the physical Hilbert space.\footnote{The reader may wonder why we do not quotient by this global symmetry.  We could, this would correspond to projecting onto the charge zero sector of the theory.  From a quantum mechanical point of view there is nothing wrong with this, but it removes states that we might be interested in studying and it most likely leads to a nonlocal theory on the boundary (this follows from the unproven lore that a CFT with a conserved current must also have operators charged under that current).}

The Wilson line in the charge $n$ representation associated to a curve $\Gamma$ is defined as
\be
W_n(\Gamma)=e^{i n \int_\Gamma A},
\ee
where by abuse of notation $A$ in the exponent indicates the pullback of $A$ to $\Gamma$.  I will often refer to the $n=1$ representation as the fundamental representation.  

We will often be thinking in the Hamiltonian formalism; to describe this in curved spacetime it is very convenient to introduce a covariant ADM decomposition of the metric via a (past-pointing) normal vector $n^\mu$, a time coordinate vector $t^\mu\equiv \delta_0^\mu$, the lapse and shift $t^\mu=-N n^\mu+N^\mu$, with $n_\mu N^\mu=0$, and the projection tensor $\gamma_{\mu\nu}\equiv g_{\mu\nu}+n_\mu n_\nu$.
The canonical momentum conjugate to $A_\mu$ is
\be
P^\mu\equiv -\sqrt{\gamma} E^\mu=q^{-2}\sqrt{\gamma}n_\nu F^{\mu\nu},
\ee
where we see that $n_\mu P^\mu=0$.  Within a surface of fixed $t$ we can pick an oriented codimension one surface $\Sigma$, of total codimension two, and then define the electric flux $Q_\Sigma$ as
\be\label{charge}
Q_\Sigma \equiv \int_\Sigma \sqrt{\hat{\gamma}}r_\mu E^\mu,
\ee
where $r_\mu$ is the ``outward'' pointing normal to $\Sigma$ and $\hat{\gamma}_{\mu\nu}$ is the projection tensor on $\Sigma$, ie $\hat{\gamma}_{\mu\nu}\equiv \gamma_{\mu\nu}-r_\mu r_\nu$. Quantum mechanically  $Q_\Sigma$ is quantized in integer units.  

In differential form notation we can simply write $Q_\Sigma=\frac{1}{q^2}\int_\Sigma *F$, but \eqref{charge} makes its noncommutativity with spatial Wilson lines that puncture $\Sigma$ manifest.  Indeed say a spatial curve $\Gamma$ punctures the surface $\Sigma$ $m_\uparrow$ times from inside to outside and $m_\downarrow$ times from outside to inside.  From the canonical commutation relations for $A_\mu$ and $P^\nu$ we have
\be\label{Walg}
[W_n(\Gamma),Q_\Sigma]=n(m_\uparrow-m_\downarrow)W_n(\Gamma).
\ee
Thus a spatial Wilson line puncturing $\Sigma$ an unbalanced number of times is either a creation or annihilation operator for some integral amount of $Q_\Sigma$.

The Hamiltonian is
\begin{align}\nonumber
H=\int d^{d-1} x\sqrt{\gamma}\Big[&N\left(\frac{q^2}{2\gamma} P^\mu P_\mu+\frac{1}{4q^2}F_{\mu\nu} F_{\alpha\beta}\gamma^{\mu\alpha}\gamma^{\nu\beta}-J^iA_i\right)+N^\mu \left(\frac{1}{\sqrt{\gamma}}P^\nu F_{\mu\nu}\right)\\
&-A_0\left(NJ^0+\frac{1}{\sqrt{\gamma}}\partial_\mu P^\mu\right)\Big].
\end{align}
The term multiplying $A_0$ is the Gauss constraint, which must vanish on any physical state.  In flat space this reduces to the usual formula
\begin{align}\nonumber
H=\int d^{d-1} x\Big[\left(\frac{q^2}{2} P_i P_i+\frac{1}{4q^2}F_{ij} F_{ij}-J^iA_i\right)-A_0\left(J^0+\partial_i P_i\right)\Big],
\end{align}
whose discretization is the starting point for Hamiltonian $U(1)$ lattice gauge theory \cite{Kogut:1974ag}.

\subsection{Reconstruction near the vacuum}
The standard CFT reconstruction of bulk fields \cite{Banks:1998dd,Hamilton:2006az,Kabat:2011rz,Heemskerk:2012mn,Morrison:2014jha} proceeds by solving the bulk equations of motion of the CFT, with boundary conditions given by the ``extrapolate'' operator dictionary of AdS/CFT \cite{Harlow:2011ke} (for a recent review of the correspondence more generally, see section six of \cite{Harlow:2014yka}).  These methods suffice to reconstruct all bulk operators acting on states in the vicinity of the vacuum, such as those shown in the left diagram of figure \ref{giopsfig}.  For example consider a free gauge field in the Poincare patch of $AdS_d$, with metric
\be
ds^2 =\frac{dz^2+\eta_{\mu\nu}dx^\mu dx^\nu}{z^2}.
\ee
I will work on ``holographic gauge'' $A_z=0$.  We can write the gauge field as
\be\label{Aeq2}
A_\mu(z,x)=\sum_\sigma \int \frac{d^{d-1}k}{(2\pi)^{d-1}}\Theta(-k^2)\theta(k^0)\Big(f_{\mu,k\sigma}(z,x)a_{k\sigma}+c.c\Big),
\ee
where $f_{\mu,k\sigma}$ are a complete set of positive-frequency modes, obeying the boundary condition \eqref{bc} as well as the gauge fixing condition.  They are given explicitly by
\be
f_{\mu,k\sigma}(z,x)\equiv \sqrt{\pi} e_\mu(k,\sigma) z^{\frac{d-3}{2}}J_{\frac{d-3}{2}}(\sqrt{-k^2}z) e^{ik\cdot x},
\ee
where ``positive frequency'' means $\omega\equiv k^0>0$, $k^2\equiv \eta_{\mu\nu}k^\mu k^\nu$, and $k\cdot x\equiv \eta_{\mu\nu}k^\mu x^\nu$.  The step functions in \eqref{Aeq2} ensure that we restrict to positive frequency solutions with $k$  timelike, since spacelike and null momenta give modes that are not normalizeable as $z\to \infty$.\footnote{``Normalizeable'' is defined with respect to the natural Klein-Gordon-like inner product on solutions of Maxwell's equation, given in the ADM notation defined above by $$(\widetilde{A},A)\equiv i\int d^{d-1}x\sqrt{\gamma}\left(\widetilde{A}_\mu ^*n_\nu F^{\mu\nu}-\widetilde{F}^{\mu\nu*}A_\mu n_\nu\right).$$}  The polarization vectors $e_\mu(k,\sigma)$ obey $e_\mu k^\mu=0$ and $e_\mu(k,\sigma) e_\nu(k,\sigma')\eta^{\mu\nu}=\delta_{\sigma\sigma'}$.  These modes are normalized so that the creation and annihilation operators in \eqref{Aeq2} obey 

\be
[a_{k\sigma},a_{k'\sigma'}]=\delta_{\sigma\sigma'}(2\pi)^{d-2}\delta^{d-1}(k-k').
\ee
Near the boundary at $z=0$, we have
\be\label{asymp}
f_{\mu,k\sigma}(x,z)\to N_{k}e_\mu(k,\sigma) z^{d-3} e^{ik\cdot x},
\ee
where 
\be
N_k\equiv \frac{(-k^2)^{\frac{d-3}{4}}}{2^{\frac{d-3}{2}}\Gamma\left(\frac{d-1}{2}\right)}
\ee

To reconstruct this gauge field in the CFT, we use the extrapolate dictionary
\be\label{extr}
\lim_{z\to 0} z^{3-d}A_\mu(z,x)\equiv \frac{1}{d-3} j_\mu(x),
\ee
where $j_\mu$ is the CFT current for the $U(1)$ global symmetry generated by the charge operator.  The normalization factor arises from insisting that the charge \eqref{charge}, evaluated at the boundary, becomes the integral of $j^0$.\footnote{This may look alarming from the point of view of what happens when $d=3$, but actually the pure Maxwell theory in $AdS_3$ is rather sick in the infrared.  In all known examples of $AdS_3/CFT_2$ there is always a Chern-Simons term in the bulk that regulates this.}   Comparing equations \eqref{Aeq2}, \eqref{asymp}, and \eqref{extr}, we see that
\be
N_k a_{k\sigma}= \frac{1}{d-3}e_\mu(k,\sigma)j^\mu(k).
\ee
Viewing this as a CFT definition of $a_{k\sigma}$, we can insert it back into \eqref{Aeq2} to get a CFT representation for $A_\mu$, of the form  \cite{Kabat:2012hp,Kabat:2012av,Heemskerk:2012np}
\be
A_\mu(z,x)=\int d^{d-1} x' K_{\mu\nu}(z,x;x')j^\nu(x').
\ee
A similar expression can also be derived for the CFT representation of any charged field in the bulk, and it is understood how to include interactions perturbatively using similar methods \cite{Kabat:2011rz,Heemskerk:2012mn}.  These holographic gauge operators always have gauge-invariant descriptions where we simply add dressing Wilson lines in the $z$ direction, extending to the boundary at $z=0$ \cite{Heemskerk:2012np}.  These lines can be confirmed by studying the commutator of these operators with the electric field.
  
\subsection{Reconstruction in the AdS-Schwarzschild background}
Let's now consider gauge-field reconstruction in the AdS-Schwarzschild geometry, shown in the right diagram of figure \ref{giopsfig}.  The geometry has the form
\be
ds^2=-f(r)dt^2+\frac{dr^2}{f(r)}+r^2 d\Omega_{d-2}^2.
\ee
Operators which lie entirely in the right exterior or the left exterior can be immediately reconstructed using the same technique as the previous subsection; we expand in a complete set of Schwarzschild modes and then match creation/annihilation operators to the Fourier modes of the boundary current $j$ (or the charged operators dual to any charged fields).  It is again easiest to work in holographic gauge $A_r=0$.  We now run into a new issue however; we are not actually able to go to this gauge everywhere on a spatial slice containing the bifurcation point, since this would remove our wormhole-threading Wilson line!  What happens instead is that there is always a single degree of freedom remaining at each point on the bifurcation surface, which is the holographic gauge version of a Wilson line threading the wormhole at that point.  These degrees of freedom cannot be reconstructed using the method just described, since in holographic gauge there is no set of exterior operators that become these ones in some limit (this distinguishes them from, for example, the electric flux through the bifurcation surface).   This is not an accident of holographic gauge: we saw in the introduction that the factorization of the two-CFT Hilbert space prevents any description in terms of the current alone.  If we \textit{can} find a description of these degrees of freedom however, such as that advocated in the main text, then as already mentioned in the introduction we can then use bulk Cauchy evolution to evolve them together with the exterior operators up into the interior \cite{Heemskerk:2012mn}.

Finally, in comparing with the CFT, it is important to remember that the thermofield double state is defined as
\be
|\psi\ran=\frac{1}{\sqrt{Z(\beta)}}\sum_i e^{-\beta E_i/2}|i^*\ran|i\ran,
\ee
where $|i\ran$ is some complete basis of energy eigenstates of the CFT quantized on the sphere and $|i^*\ran=\Theta|i\ran$, where $\Theta$ denotes the natural CPT-like operation that exchanges the two CFTs and reverses time.  This $\Theta$ operation is the reason that it is $Q_R+Q_L$, rather than $Q_R-Q_L$, that annihilates the thermofield double state.    Charge fluctuations are correlated between the two sides in such a way that the total charge is zero.  Here $Q_R$ and $Q_L$ are both defined as the integral of $j^0$ over the relevant boundary.  From \eqref{Walg}, we see that a wormhole-threading Wilson line will commute with $Q_R+Q_L$, as it must since it does not create charge in the bulk, but act as a creation or annihilation operator for $2n$ units of $Q_R-Q_L$.

\bibliographystyle{jhep}
\bibliography{bibliography}
\end{document}